\documentclass[preprint,superscriptaddress,amsmath,amssymb,pra]{revtex4}
\usepackage{graphicx}
\usepackage{subfigure}
\usepackage{graphicx}
\usepackage{dcolumn}
\usepackage{bm}
\usepackage{amsmath}
\usepackage{setspace}
\usepackage{mathrsfs}

\newcommand{\anbr}[1]{\left\langle #1 \right\rangle}
\newcommand{\DvR}[1]{\left| #1 \right\rangle}
\newcommand{\DvL}[1]{\left\langle #1 \right|}
\newcommand{\DvIP}[2]{\left\langle #1 | #2 \right\rangle}
\newcommand{\ident}[0]{\openone}%
\newcommand{\hmlt}[0]{\mathscr{H}}

\begin{document}

\title{Signatures of incoherence in a quantum information processor}

\author{Michael K. Henry}
\affiliation{Department of Nuclear Science and Engineering, Massachusetts Institute of Technology, Cambridge, Massachusetts 02139, USA}
\author{Alexey V. Gorshkov}
\affiliation{Physics Department, Harvard University, Cambridge, Massachusetts 02138, USA}
\author{Yaakov S. Weinstein}
\affiliation{Quantum Information Science Group, The {\sc Mitre} Corporation, Eatontown, New Jersey 07724, USA}
\author{Paola Cappellaro}
\affiliation{Department of Nuclear Science and Engineering, Massachusetts Institute of Technology, Cambridge, Massachusetts 02139, USA}
\author{Joseph Emerson}
\affiliation{Department of Applied Mathematics and Institute for Quantum Computing, University of Waterloo, Waterloo, Ontario N2L 3G1, Canada}
\author{Nicolas Boulant}
\affiliation{Department of Nuclear Science and Engineering, Massachusetts Institute of Technology, Cambridge, Massachusetts 02139, USA}
\author{Jonathan S. Hodges}
\affiliation{Department of Nuclear Science and Engineering, Massachusetts Institute of Technology, Cambridge, Massachusetts 02139, USA}
\author{Chandrasekhar Ramanathan}
\affiliation{Department of Nuclear Science and Engineering, Massachusetts Institute of Technology, Cambridge, Massachusetts 02139, USA}
\author{Timothy F. Havel}
\affiliation{Department of Nuclear Science and Engineering, Massachusetts Institute of Technology, Cambridge, Massachusetts 02139, USA}
\author{Rudy Martinez}
\affiliation{Department of Natural Sciences, New Mexico Highlands University, Las Vegas, New Mexico 87701, USA}
\author{David G. Cory}
\affiliation{Department of Nuclear Science and Engineering, Massachusetts Institute of Technology, Cambridge, Massachusetts 02139, USA}

\date{\today}

\begin{abstract}
Incoherent noise is manifest in measurements of expectation values when the underlying ensemble evolves under a classical distribution of unitary processes.  While many incoherent processes appear decoherent, there are important differences.  The distribution functions underlying incoherent processes are either static or slowly varying with respect to control operations and so the errors introduced by these distributions are refocusable.  The observation and control of incoherence in small Hilbert spaces is well known.  Here we explore incoherence during an entangling operation, such as is relevant in quantum information processing.  As expected, it is more difficult to separate incoherence and decoherence over such processes. However, by studying the fidelity decay under a cyclic entangling map we are able to identify distinctive experimental signatures of incoherence.  This is demonstrated both through numerical simulations and experimentally in a three qubit nuclear magnetic resonance implementation.
\end{abstract}

\maketitle

Incoherent noise encodes quantum information into the classical degrees of freedom of an ensemble by a distribution of unitary errors \cite{Pravia2003,Boulant2004}.  An incoherent process is tied to a time-independent or slowly varying classical probability distribution of Hamiltonians.  Evolution under such a process is naturally described as an operator sum, given in superoperator notation \cite{Havel2003, Weinstein2004} by
\begin{align}
\label{ch3eq:supincohsum}
\hat{S}(t)=\int p(z) e^{i \hmlt^*(z) t} \otimes e^{-i \hmlt(z) t} dz
\end{align}
where $p(z)$ is the classical probability distribution of Hamiltonians $\hmlt(z)$ (* denotes complex conjugation).  A variety of tools have been developed to counteract incoherent noise in quantum information processors (QIPs). Optimal control theory minimizes the errors caused by uncertainty in the system Hamiltonian \cite{Dahleh1990,Peirce1988}.  Dynamical decoupling and bang-bang control actively suppress incoherence by periodically refocusing part of the evolution \cite{Viola1998,Viola1999a,Viola1999b,Viola2002}. Strongly modulating pulses \cite{Pravia2003} and composite pulses \cite{Levitt1986} have also been used to refocus incoherent noise.  Such techniques exploit the reversibility of incoherent errors and are particularly valuable since they do not require access to a larger Hilbert space, as do decoherence free subspaces \cite{Zanardi1997,Lidar1998,Duan1997}, noiseless subsystems \cite{Knill2000,Zanardi1997} and other quantum error encodings \cite{Shor1995,Steane1996}.  Decoherent noise by contrast does require the full power of quantum error-correcting codes, so distinguishing the presence of incoherence is important in choosing an error-correction scheme.

Incoherence, which is typically studied for single-qubit errors in SU(2), causes a loss of purity in the ensemble-averaged state while preserving the purity of the individual ensemble members.  Decoherence is a distinct process that irreversibly reduces the purity of the individual ensemble members.  In small Hilbert spaces, incoherence is easily detected and controlled either by time reversal of the control field or through creation of echoes.  Some classic examples include the rotary echo \cite{Rhim1971a,Kessemeier1972}, the Hahn echo \cite{Hahn1950}, and the Carr-Purcell and CPMG echo sequences \cite{Carr1954, Meiboom1958}.  In these examples, incoherent errors are completely refocused by an inverted incoherent process, and the resulting increase in purity over the ensemble causes an observable echo.  Identifying and controlling incoherence is more difficult in Hilbert spaces that support entanglement and in particular, in the presence of an entangling operation.  An entangling operation propagates incoherent errors to non-separable states, causing a loss of purity that is not recovered by an inverted incoherent process, so the incoherence mimics a decoherent process.

Here we present an example of incoherence influencing an entangling operation in a three-qubit liquid state nuclear magnetic resonance (NMR) QIP, and we show how the incoherence appears as a distinct process from decoherence in the measurement of fidelity decay under imperfect time reversal.  Fidelity decay \cite{Peres1984, Jacquod2001} has previously been shown to be a useful tool for efficiently characterizing errors in a QIP \cite{Emerson2002}.  In the method suggested here, the task of measuring fidelity decay is simplified by studying fidelity decay under a cyclic operation, which removes the need to invert the ideal evolution and admits analysis by Average Hamiltonian Theory \cite{Haeberlen1968}.  We show that in our experiment incoherence causes recurrences in fidelity that could not arise from a decoherent process satisfying certain well-defined properties, in particular those of a Markovian semigroup, which are known to apply in a broad set of conditions \cite{Alicki1987}.  The signature of incoherence observed in experimental data is also analyzed by numerical simulations of the NMR experiment.

\section{Identifying incoherence by fidelity decay}

The fidelity between two quantum states $\rho$ and $\tilde{\rho}$ is defined as the Hilbert-Schmidt inner product
\begin{align}
F = \DvIP{\tilde{\rho}}{\rho} = \text{trace}\left( \tilde{\rho}^\dagger\rho \right),
\end{align}
where $\DvR{\rho}$ is the density matrix represented as a state vector in Liouville space.  Given an ideal unitary map described by the superoperator $\hat{S}$, a perturbation described by the superoperator $\hat{P}$ and an initial state $\rho_0$, the fidelity decay after $n$ iterations of imperfect time reversal is 
\begin{align}
F_n = \DvL{\rho_0} \left(\hat{S}^{-1}\right)^n \left(\hat{P}\hat{S}\right)^n \DvR{\rho_0}.
\end{align}
Here we consider the case where $\left(\hat{P}\hat{S}\right)$ is a noisy implementation of $\hat{S}$, and therefore implementing the ideal inverse map $\hat{S}^{-1}$ is impractical or at best inconvenient.  However, if we choose $\hat{S}$ to be cyclic, then for some number of iterations $n_c$, we have $(\hat{S}^{-1})^{n_c} = \hat{S}^{n_c}=\ident$.  Now if we constrain the fidelity decay to be measured only after iterations that are an integer multiple of $n_c$, we have an expression in the form of the survival probability
\begin{align}
F_n = \DvL{\rho_0} \left(\hat{P}\hat{S}\right)^{n\cdot n_c} \DvR{\rho_0},
\end{align}
thus simplifying the fidelity decay expression and measurement.  There are a number of cyclic superoperators that are useful for quantum information processing, for example, the Hadamard gate, the controlled-NOT gate, the swap gate, and the quantum Fourier transform.  In the experiment described here, we measure the fidelity of a cyclic entangling operation where $n_c=8$.  In addition to measurements after each full cycle, we measure the fidelity after each half cycle (every fourth iteration), since the ideal output at these time points is also a trivially known separable state.

The algorithmic efficiency of measuring fidelity decay \cite{Emerson2002} makes it an attractive tool for characterizing errors in a QIP. In our experiment, we are interested in using fidelity decay to distinguish incoherent and decoherent noise processes.  We will see that the presence of recurrences (or periodic increases) in the fidelity decay is a signature of incoherent noise.  First we describe a framework for discussing noise and measurement in an ensemble QIP.  

Let $\hat{S}$ and $\hat{P}_z$ (superoperator matrices of size $N^2\times N^2$) represent unitary processes over a Hilbert space of dimension $N$, where $z$ is a classical parameter that labels the ensemble members. $\hat{P}_z$ is a perturbation of the form $\exp{(-i\eta_z\hat{V}_z)}$, where $\hat{V}_z$ is an Hermitian operator in the $N$ dimensional Hilbert space and $\eta_z$ is the strength of the noise for a particular member of the ensemble.  The ideal map $\hat{S}$ acting $n$ times on the initial state $\rho_0$ returns the state $\hat{S}^n\DvR{\rho_0}=\DvR{\rho_n}$, while the perturbed map returns the state $(\hat{P}_z\hat{S})^n\DvR{\rho_0}= \DvR{\tilde{\rho}_n(z)}$. The input state and the ideal output state have no dependence on the classical parameter $z$.  The perturbation and (consequently) the perturbed output state both have an explicit $z$-dependence.  In an experiment, we measure the ensemble-averaged state $\anbr{\DvR{\tilde{\rho}_n(z)}}_z=\DvR{\tilde{\rho}_z}$, where $\anbr{\cdot}_z\Rightarrow\int(\cdot)p(z)dz$, and $p(z)$ is the probability of measuring the ensemble member labeled $z$.  In an NMR QIP for example, $p(z)$ may represent the physical fraction of the ensemble associated with a particular value of the radio frequency (rf) control field strength.

Incoherence, as previously explained, describes a process whereby information is reversibly encoded in the classical degrees of freedom of an ensemble by a static or slowly varying distribution of Hamiltonians.  Incoherent dynamics are modeled by considering $\hat{P}_z$ to be a static perturbation for each member of the ensemble over all iterations of the map.  In this case, the output state is given by $\DvR{\tilde{\rho}_n}=\anbr{\left(\hat{P}_z\hat{S}\right)^n\DvR{\rho_0}}_z$, and the corresponding fidelity after $n$ iterations is 
\begin{equation}
F_n = \DvIP{\rho_n}{\tilde{\rho}_n} = \DvL{\rho_n}\anbr{\left( \hat{P}_z\hat{S} \right)^n\DvR{\rho_0}}_z.
\end{equation}
Incoherence causes a loss of purity in the ensemble-averaged state, but the purity of the individual ensemble members is preserved since the local system dynamics are unitary.  Consequently, small errors which initially cause a decrease in fidelity can in principle be ``refocused'' by later iterations of the map, resulting in fidelity recurrences.  

In contrast to incoherence, decoherence describes a process whereby information is irreversibly lost to an environment.  Decoherence is often modeled as a coupling to an expanded Hilbert space that includes an unobservable environment.  The instantaneous state of the system is found by taking a partial trace over the environmental degrees of freedom, thus removing any information that has ``leaked'' into the environment.  Markovian decoherence, which we consider here, can be modeled semi-classically as a stochastic process \cite{AbragamTextCh8}.  In this description, evolution of the quantum state is modeled by averaging the system dynamics over a random distribution of unitary processes.  Here we consider the case where $\hat{P}_z$ and $p(z)$ describe the random distribution of unitary processes, and an identical decoherent noise process is implemented upon each iteration of the map. The output state in this case is given by $\DvR{\tilde{\rho}_n}=\anbr{\left(\hat{P}_z\hat{S}\right)}_z^n\DvR{\rho_0}$, and the corresponding fidelity after $n$ iterations is 
\begin{equation}
F_n = \DvL{\rho_n}\anbr{\left( \hat{P}_z\hat{S} \right)}_z^n\DvR{\rho_0}.
\end{equation}
The individual ensemble members (and thus the ensemble-averaged state as well) lose purity under a decoherent process, so fidelity losses due to Markovian decoherence are not refocused by further iterations of the map.  The resulting fidelity decay decreases exponentially before saturating at $1/N$, a well-known result for decoherent noise \cite{Jacquod2001}.  The possibility of fully refocusing errors in the incoherent case is the essential difference between decoherent and incoherent dynamics, and this difference is what leads to observable signatures of incoherence.  

There is also a third type of noise that is often discussed for QIPs: coherent noise causes non-ideal unitary errors that are uniform over the ensemble and do not cause a loss of purity in the individual ensemble members or in the ensemble-averaged state.  Like incoherence, coherent noise can cause recurrences in fidelity decay.  However, there is little motivation to distinguish these two noise processes in the setting of quantum information processing since they both can be treated with the same techniques, which do not require access to a larger Hilbert space.

Recurrences in a fidelity decay are a signature of microscopically reversible dynamics. The recurrence in fidelity decay is more general than a simple echo experiment in which incoherence is inverted by local SU(2) operations.  Recurrence in fidelity decay can result from errors refocused from any part of Hilbert space through the repeated action of the perturbed map.  For the case that $\hat{S}$ is an entangling operation, incoherent errors will cause a loss in the purity of the ensemble that is not recovered by single-qubit operations, and therefore is difficult to distinguish from the effects of decoherence. Fidelity decay under imperfect time reversal provides an efficient means for observing signatures of incoherence even in the presence of an entangling operation.

\section{Experiment}  

\subsection{Quantum Circuit}
\begin{figure}
	\begin{center}
 	\includegraphics[width=3in]{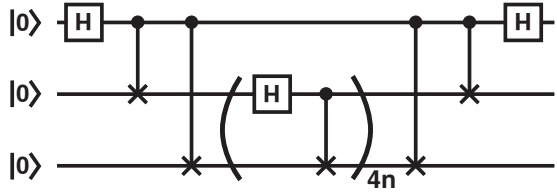}
 	\end{center}
	\caption{\label{ch3fig:circuit} The quantum circuit for exploring incoherence in an entangling operation on a QIP.  H represents the single-qubit Hadamard operation $(\sigma_z+\sigma_x)/\sqrt{2}$, the two-qubit gate represents a controlled-NOT operation, which flips the target qubit when the control qubit is in the $\DvR{1}$ state.  The first three gates create a maximally entangled GHZ state $\left( \DvR{000} +\DvR{111} \right)/\sqrt{2}$, which is followed by $4n$ iterations of a two-qubit entangling operation.  The final three gates convert the resulting entangled state to the computational basis state $\DvR{000}$ for even values of $n$ and $\DvR{001}$ for odd values of $n$.  Incoherence in an entangling operation mimics decoherence by causing a loss of purity that is not refocused in the output state by inverting the incoherence.  In our experiment, we observe signatures of incoherence in the two-qubit entangling operation by measuring the fidelity decay of the output state for $n$=0 through 30.}
\end{figure}

Figure \ref{ch3fig:circuit} shows the quantum circuit used to study incoherence in an entangling operation.  
The first three gates in the circuit create the GHZ state $\left( \DvR{000} +\DvR{111} \right)/\sqrt{2}$.  Next, an entangling operation on qubits two and three is repeated $4n$ times.  The final three gates convert the resulting entangled state to a computational basis state.  For odd values of $n$ the final state is $\DvR{001}$, and for even values of $n$ the final state is $\DvR{000}$.  For all integer values of $n$, the ideal output state is separable (i.e. non-entangled).  Incoherent noise in the iterated portion of the circuit will create entanglement in the observed output state, thus attenuating the purity of the reduced states of individual qubits as well as the purity of the overall three-qubit quantum state.  Because entanglement cannot be removed by single qubit operations, the purity losses caused by incoherence cannot be reversed by inverting the incoherence on the output state.  The circuit in Fig. \ref{ch3fig:circuit} is an example of a case where incoherence causes errors in the output state that are not easily distinguished from the effects of decoherence.  
  
\subsection{NMR QIP}
\begin{figure}
	\begin{center}
	\includegraphics[width=3in]{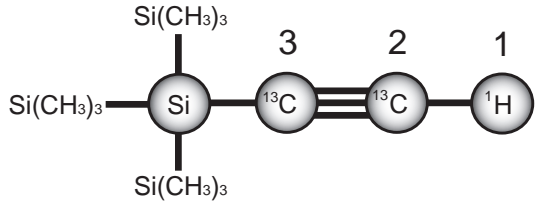}
 	\end{center}
	\caption {A diagram of the tris(trimethylsilyl)silane-acetylene molecule used to implement the quantum circuit in Fig. \ref{ch3fig:circuit} in a liquid state NMR QIP.  The two-qubit entangling operation is applied to the two $^{13}$C spins in the acetylene branch, labeled qubits 2 and 3.  The primary source of incoherence in the experimentally implemented entangling operation is the inhomogeneity of the carbon rf control field.}
	\label{ch3fig:TMSA}
\end{figure}

In an NMR QIP, nuclear spins polarized by a strong external magnetic field serve as qubits.  The molecule used in this experiment, diagrammed in Fig. \ref{ch3fig:TMSA}, is tris(trimethylsilyl)silane-acetylene dissolved in deuterated chloroform (250 millimolar solution).
The carbon nuclei in the acetylene branch are isotopically enriched $^{13}$C, while the methyl carbons are of natural isotopic abundance.  The two carbon-13 nuclei and the hydrogen nucleus in the acetylene branch are used as qubits.  The full internal Hamiltonian of the nuclear spin system has the form
\begin{equation}
\hmlt_{int}=\sum_{j=1}^{n_q}{\pi\nu_j\sigma_z^j}+\sum_{j<k}{\frac{\pi J_{jk}}{2}\sigma^j\cdot\sigma^k}
\end{equation}
where $\nu_j$ is the resonance frequency of the $j^{th}$ spin, $J_{jk}$ is the frequency of scalar coupling between spins $j$ and $k$, and $\sigma^j$ represents a generalized Pauli operator acting on the $j^{th}$ spin.  The hydrogen nucleus is labeled qubit number $1$, making it the most significant bit in the computational state vector.  The repeated entangling operation is applied to the carbon qubits, which are labeled as indicated in Fig. \ref{ch3fig:TMSA}.  Experiments are performed in a 9.4 Tesla magnetic field at temperature 300 K, where the Carbon qubits are separated by 1.201 kHz.  The scalar couplings are $J_{12}=235.7$ Hz, $J_{23}=132.6$ Hz, and $J_{13}=42.9$ Hz.  

The input pseudopure state \cite{Cory1997} was created by the technique described in \cite{Teklemariam2001} using hard rf pulses and gradient fields.  The input pseudopure state preparation pulse sequence, which is non-unitary, was optimized based on the state correlation \cite{Fortunato2002} between the numerically simulated input state and the ideal input state.  The average gate fidelities \cite{Fortunato2002} of the sequences corresponding to the three sections of the circuit were optimized over the full Hilbert space.  In the experiment, representative measurements of the fidelity are taken.  A single $\pi/2$ readout pulse on the hydrogen spin converts the $\sigma_z^1$, $\sigma_z^1\sigma_z^2$, $\sigma_z^1\sigma_z^3$, and $\sigma_z^1\sigma_z^2\sigma_z^3$ components of the output density matrix to observable signal, for $n=0$ through $n=30$.  

\section{Numerical Simulation}

\begin{figure}
	\begin{center}
 	\includegraphics[width=3in]{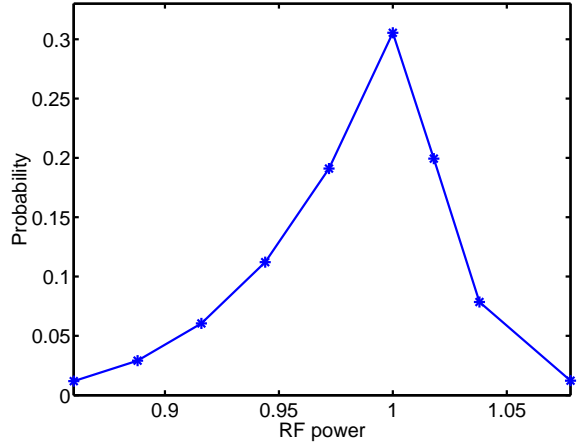}
 	\end{center}
	\caption{The distribution of carbon rf powers measured in previous experiments and used in numerical simulations of the NMR implementation of the circuit in Fig. \ref{ch3fig:circuit}.  The rf power is in units of the nominal rf nutation frequency (17.5 kHz).}
	\label{ch3fig:rf_pdf}
\end{figure}

Numerical simulations demonstrate the signatures of incoherence that we expect to see in the NMR experiment. The dominant incoherent noise in the experiment arises due to the inhomogeneity of the rf control field over the spatial extent of the liquid state NMR sample.  When an rf pulse is applied during the experiment, the members of the ensemble experience a distribution of rf powers, and only a fraction of the ensemble actually experience precisely the nominal (ideal) rf power.   While the control fields for both the hydrogen and carbon qubits are known to be inhomogeneous, the inhomogeneity of the carbon control field is the dominant source of incoherent errors in the entangling operation.  Consequently, our numerical simulations include incoherence for each carbon pulse as a distribution of rf control field strengths.  The discrete nine-point distribution measured in previous experiments and used in simulations is plotted in Fig. \ref{ch3fig:rf_pdf}. 
The natural decoherence of the nuclear spin system is simulated by an approximate relaxation superoperator \cite{ErnstText}, which is completely diagonal in the generalized Pauli basis.  In this diagonal form, each non-zero entry in the relaxation superoperator represents the decoherence rate of a generalized Pauli basis operator; the specific values used in simulations are based on measurements of all T$_1$s (ranging from 3.0 to 10.4 seconds) and the single species T$_2$s (ranging from 1.5 to 3.0 seconds).  Four cycles of the Hadamard Controlled-NOT sequence are implemented in a pulse sequence lasting 34 milliseconds. 

In numerical simulations, we are interested in the unique features of fidelity decay caused by incoherence.  We isolate the effects of incoherence by simulating the rf inhomogeneity in two regimes of dynamics.  In the incoherent model, rf inhomogeneity is simulated as it actually occurs in the experiment, as a static distribution of local unitary noise.  The output state in this regime is $\anbr{\left(\hat{P}_z\hat{S}\right)^n\DvR{\rho_0}}_z$, where $z$ now represents the power of the carbon rf control field. In this model, the pure state of each individual ensemble member is carried over to the input of the next iteration.  By contrast, in the decoherent model rf inhomogeneity is simulated fictitiously as a stochastic process (having zero correlation time). The output state in this regime is $\left(\anbr{\hat{P}_z\hat{S}}_z\right)^n\DvR{\rho_0}$.  In the decoherent model, the purity of the individual ensemble members is attenuated and only the average state of the ensemble is carried over to the input of the next iteration.  

Differences between the two models arise purely from the manner in which rf inhomogeneity is simulated - the incoherent model is a deterministic process for each ensemble member while the decoherent model is a stochastic process for each ensemble member.  We emphasize that rf inhomogeneity is known to be a deterministic process on the time scales of our experiment, and the fictitious stochastic model is used only to isolate the signatures of incoherence.  We also note that the relaxation superoperator, which represents well-understood naturally-occurring decoherent noise that occurs in the experiment, is simulated identically in both models.  The only difference between the two models is the manner in which rf inhomogeneity is simulated.

\begin{figure}
	\begin{center}
	\includegraphics[width=3.2in]{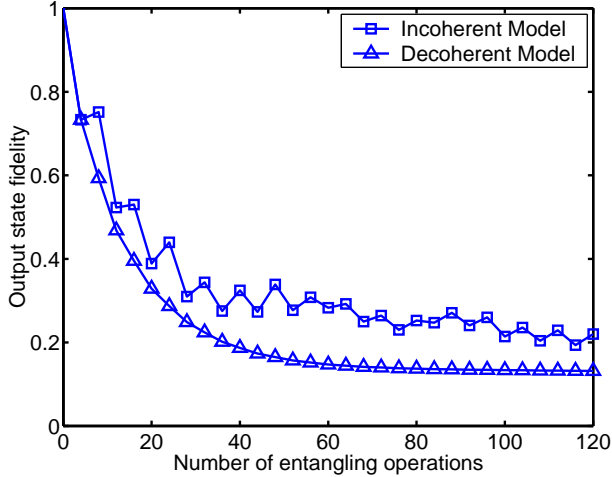}
 	\end{center}
	\caption{  The fidelity decay from a numerical simulation of the experiment, where rf inhomogeneity (RFI) is simulated using two different models.  Data points are connected by straight lines to guide the eye.  RFI is simulated as a static distribution in the incoherent model (squares), while RFI is fictitiously simulated as a stochastic effect in the decoherent model (triangles).  This fiction allows us to isolate signatures of incoherence.  In the incoherent model, fidelity recurrences (which appear as oscillations in the plot) are observed because the purity of the ensemble members is preserved and the repeated action of the entangling map refocuses some of the errors.  There are no significant recurrences in the decoherent model because the individual members of the ensemble lose purity, the errors are not refocused, and the fidelity decays steadily and saturates at $1/N = 1/8$.  Comparison of the two plots shows that the fidelity decay recurrences are caused by incoherent noise.   }
	\label{ch3fig:fidecay}
\end{figure}

The results of numerical simulations are plotted in Fig. \ref{ch3fig:fidecay}.
Although incoherence in entangling operations creates a loss of purity that mimics decoherence, fidelity decay under imperfect reversal of such a process reveals distinguishable properties of the incoherence.  The $n=1$ point in the two fidelity decays are identical, as expected.  However, differences in the two models are manifest already in the fidelity decay at $n=2$, as the fidelity increases only in the incoherent model.  Over 120 entangling operations, the numerically simulated fidelity decay for the incoherent case shows periodic increases in fidelity, or recurrences, which are only possible for microscopically reversible dynamics.  The decoherent simulation shows a continuous decay and saturation at a value of the inverse of the dimension of the Hilbert space $1/N$.  Differences between the fidelity decays collected in the two regimes reveal a signature of incoherent noise which is also observed in experimental data.  

\section{Experimental Results and Discussion}

\begin{figure}
	\begin{center}
 	\includegraphics[width=3.2in]{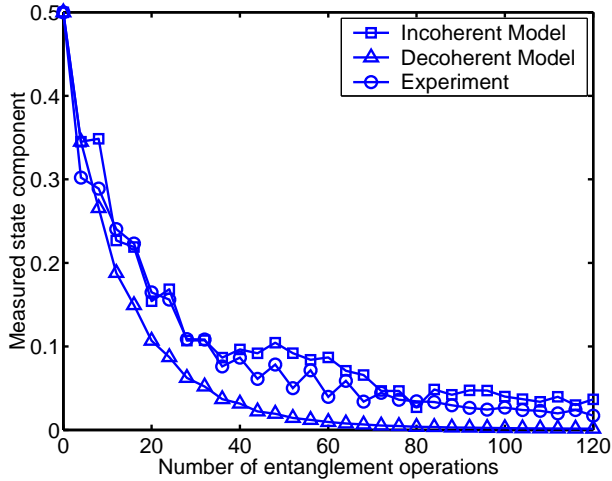}
 	\end{center}
	\caption{ The sum of the absolute value of the density matrix components measured in the experiment (circles) and in numerical simulations.  This measurement is a representative measure of state fidelity for the map under consideration.  Rf inhomogeneity is numerically simulated using an incoherent model (squares) and a fictitious decoherent model (triangles) as discussed in the text.  Data points are connected by straight lines to guide the eye.  For the ideal map with no noise, the sum of the density matrix components is a constant value of 0.5.  Experimental noise causes the measured value to initially decrease.  However, as the map is iterated, the measured value increases periodically.  This behavior is well-reproduced by the incoherent model, while the decoherent model does not predict fidelity recurrences.  This plot shows that incoherence in the entangling operation appears with distinct signatures in the experimental data.  }
	\label{ch3fig:pos_sum}
\end{figure}

Experimental data resulting from an implementation of the optimized control sequences are compared to results of numerical simulations of the experiment for the two models of rf inhomogeneity previously discussed.  Figure \ref{ch3fig:pos_sum} shows the sum of the measured magnitudes of four state components ($\sigma_z^1$, $\sigma_z^1\sigma_z^2$, $\sigma_z^1\sigma_z^3$, and $\sigma_z^1\sigma_z^2\sigma_z^3$) obtained by experiment and by numerical simulations.  
Under the ideal unitary evolution, the value of the plotted sum is 0.5.  The experimentally observed value decreases initially, shows periodic recurrences, and later becomes nearly constant.  The incoherent model reproduces the important features of the experimental data, showing significant fidelity recurrences before settling to a nearly constant value.  By contrast the value simulated by the decoherent model decreases rapidly and steadily, never increases significantly, and saturates to zero.  This comparison demonstrates that incoherence in the entangling operation appears with distinct signatures in the experimental data.  

\begin{figure*}[htb]
	\begin{center}
 	\includegraphics[width=6in]{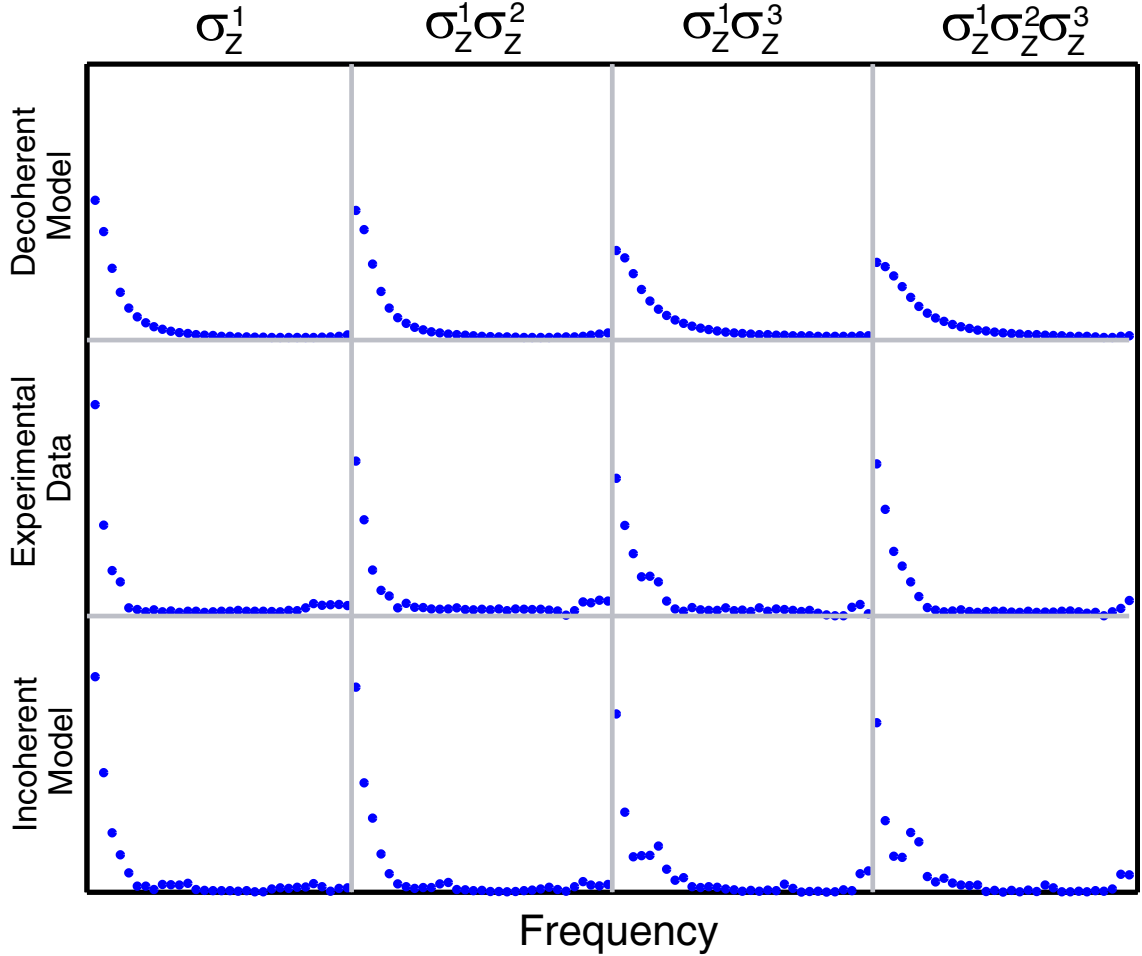}
 	\end{center}
	\caption{ The Fourier transform of each experimentally measured component of the density matrix (middle row), compared to numerical simulations of the experiment using two models of rf inhomogeneity discussed in the text. The horizontal axis represents frequency in units of oscillation periods per entangling operation, with values ranging from 0 (left) to 1/8 (right).  Resolved high frequency components, which represent fidelity recurrences, are observed in the experimental data and in numerical simulations of the incoherent model, but not in numerical simulations of the decoherent model. The dominant high frequency components are observed in the three-body term $\sigma_z^1\sigma_z^2\sigma_z^3$ at the Nyquist frequency.  This plot shows that incoherence in the experimentally implemented entangling operation appears as high frequency components in the Fourier transform of a state fidelity measurement.}
	\label{ch3fig:pos_ft}
\end{figure*}

Some insight is gained by comparing the individually measured state components of the density matrix ($\sigma_z^1$, $\sigma_z^1\sigma_z^2$, $\sigma_z^1\sigma_z^3$, and $\sigma_z^1\sigma_z^2\sigma_z^3$) in the frequency domain by Fourier transforming the data, as plotted in Fig. \ref{ch3fig:pos_ft}.  In each set of axes, the frequency is represented on the horizontal axis in units of oscillation periods per entangling operation.  The highest observable (Nyquist) frequency is 1/8, since the state was measured after every four entangling operations.
Comparing the experimental data with the two types of simulation, we see again that the incoherent model of rf inhomogeneity accurately reproduces key features of the experimental data which are not reproduced by the decoherent model.  The dominant signal in all twelve plots is the zero frequency peak, which is caused by the initial decline in fidelity observed for all three plots in Fig. \ref{ch3fig:pos_sum}.  The zero frequency peak is somewhat broader in the decoherent model, which reflects the rapid decay to zero of that data in the time-domain.  The important features in the time-domain data, namely the oscillatory fidelity recurrences, are represented in the high frequencies of the Fourier domain.  The fidelity recurrences in the experimental data and in the incoherent model simulations in Fig. \ref{ch3fig:pos_sum} appear as resolved high frequency components of the individual state measurements in Fig. \ref{ch3fig:pos_ft}.  The largest high frequency component occurs in the $\sigma_z^1\sigma_z^2\sigma_z^3$ measurement at the Nyquist frequency.

\section{Conclusions}

Incoherence in an entangling operation causes a loss of purity over the ensemble that is not recovered by an inverted incoherent process, and therefore is difficult to distinguish from decoherence.  However, incoherence  due to inhomogeneity in the rf control field during the implemented entangling operation appears as a distinct process in our experimental data in the form of fidelity recurrences.  Numerical simulations identified the recurrences as a purely incoherent effect.  Incoherent errors are isolated in numerical simulations by separating out those parts of the evolution that are identical over the ensemble in a fictitious decoherent model, and we see that the decoherent process does not give rise to fidelity recurrences. 

We have shown that incoherence can lead to recurrences in fidelity decay under a cyclic operation, and this provides an efficient benchmark for distinguishing incoherent noise from purely Markovian decoherence. In our experiment, a two-qubit entangling operation was repeated 120 times on a three-qubit GHZ state in a liquid state NMR QIP, and fidelity recurrences in the experimental data were created by incoherence due to inhomogeneity of the rf control field. The experiment was numerically simulated by modeling rf inhomogeneity in two regimes: as a static distribution of Hamiltonians, and fictitiously as a stochastic distribution of Hamiltonians. The stochastic model mimics a decoherent process, allowing us to isolate the incoherent effects of rf inhomogeneity. The comparison identifies the experimentally observed recurrences as an incoherent process. The approach for detecting incoherence described here will be a valuable resource in QIPs operating in larger Hilbert spaces with
entangled states over many qubits, where the effects of incoherence and decoherence are difficult yet important to distinguish.

\bibliography{SignaturesOfIncoherence_v2}
\end{document}